# High-$T_c$ Phase of PrO$_{0.5}$F$_{0.5}$BiS$_2$ single crystal induced by uniaxial pressure

Masaya Fujioka[1]*, Masanori Nagao[2], Saleem J. Denholme[1], Masashi Tanaka[1] Hiroyuki Takeya[1], Takahide Yamaguchi[1], and Yoshihiko Takano[1,3]

[1] National Institute for Materials Science, 1-2-1 Sengen, Tsukuba, Ibaraki 305-0047, Japan

[2] Center for Crystal Science and Technology, University of Yamanashi, Kofu 400-8511, Japan

[3] University of Tsukuba, 1-1-1Tennodai, Tsukuba, Ibaraki 305-0001, Japan

We demonstrated a pressure-induced phase transition from a low-$T_c$ phase (3.5 K) to a high-$T_c$ phase (8.7 K) in single crystalline PrO$_{0.5}$F$_{0.5}$BiS$_2$. The high-$T_c$ phase is easily observed even at 0.7 GPa by uniaxial pressure obtained from a geometric effect of a single crystal with an extremely-thin thickness. For this phase transition, superconducting anisotropy and the coherence length along the $c$ axis change from $\gamma$ = 20 to 9.3 and from $\xi_c$ = 0.56 to 0.71 nm respectively. It is suggested that a shrinkage in the $c$ lattice constant by applied uniaxial pressure enhances the three dimensionality in the superconducting state.

## Introduction

Since the discovery of the novel superconductor Bi$_4$O$_4$S$_3$ [1], various BiS$_2$-based superconductors such as LnO$_{1-x}$F$_x$BiS$_2$ (Ln = La, Ce, Pr, Nd, Yb) [2-8] and Sr$_{1-x}$F$_x$BiS$_2$ [9,10] have been reported one after the other. This family has a crystal structure composed of an alternate stacking of conducting layers and blocking layers, in common with the cuprate superconductors [11] and Fe-based superconductors [12].

LaO$_{1-x}$F$_x$BiS$_2$ is a typical BiS$_2$-based superconductor. The parent material LaOBiS$_2$ shows semiconducting behavior. It is necessary to introduce electron carriers within the BiS$_2$ layer to induce superconductivity. The partial substitution of O by F is useful for this carrier doping. Although, optimal fluorine concentration is estimated to be around x = 0.5, the obtained superconducting transition temperature ($T_c$) is completely different between conventional solid state reactions and high pressure synthesis [2, 13-15]. In the case of the conventional solid state reaction, the $T_c$ of LaO$_{0.5}$F$_{0.5}$BiS$_2$ is obtained at only 3 K. On the other hand, the $T_c$ largely increases above 10 K by using high pressure synthesis. Furthermore, even if the sample is prepared by using a conventional solid state reaction method, a high pressure measurement also enhances the $T_c$ above 10 K [16]. It is notable that, the $T_c$ does not continuously increase; in fact, it suddenly increases from a low $T_c$ around 3 K to a high $T_c$ around 10 K when applying pressure.

In recent research, Tomita *et al.* reported that this pressure-induced phase transition from low-$T_c$ to high-$T_c$ phase is caused by a structural transition from a tetragonal phase (P4/nmm) to a monoclinic phase (P2$_1$/m), according to the high-pressure X-ray diffraction measurements [16].

Previously, polycrystalline PrO$_{0.5}$F$_{0.5}$BiS$_2$ was also investigated using high-pressure measurement techniques [17, 18]. $T_c$ suddenly increases from 3.5 to 7.6 K under hydrostatic pressure from ambient pressure to 2.5 GPa. On the other hand, the pressure induced superconductivity of BiS$_2$ system has not yet been reported in single crystals. Therefore, the high-pressure measurement of a single crystal is required to investigate the intrinsic properties of this phase transition. Recently, Nagao et al. developed a method to grow single crystals of BiS$_2$-based superconductors by using CsCl/KCl flux

[19]. In this research, we have succeeded in preparing this single crystals of PrO$_{0.5}$F$_{0.5}$BiS$_2$, and applied uniaxial pressure along the *c* axis. This is the first report on the anisotropic superconducting properties of a BiS$_2$ system from ambient pressure to 2.3 GPa.

Experimental procedure

PrO$_{0.5}$F$_{0.5}$BiS$_2$ was prepared by the CsCl/KCl flux method. The raw materials (Pr$_2$S$_3$, Bi, Bi$_2$S$_3$, Bi$_2$O$_3$, BiF$_3$) were weighed at a nominal composition for PrO$_{0.5}$F$_{0.5}$BiS$_2$. They were mixed with CsCl and KCl in a mortar, and sealed in an evacuated quartz tube. The weight of the raw materials and CsCl/KCl flux were 0.8 and 3.0 g respectively, and the molar ratio of the flux was CsCl : KCl = 5 : 3. These samples were heated at 800 °C for 10 h, then slowly cooled down to 600 °C at a rate of 0.5 °C /h, and then furnace-cooled down to room temperature. Obtained samples were submerged in the distilled water to remove the flux.

X-ray diffraction (XRD) using Cu Kα radiation was measured for the characterization of the obtained samples. The surface of the sample was observed by scanning electron microscope (SEM). The elemental compositions of the samples were investigated by energy dispersive X-ray spectrometry (EDX). Electrical resistivity was measured by a physical properties measurement system (PPMS) under varying pressures up to a maximum of 2.3 GPa. The measurement was performed by a standard four-probe method from 300 K to 2 K in a piston-cylinder-type pressure cell. The measured sample was formed into rectangles with dimensions of around 1.8 × 0.6 × 0.03 mm$^2$. Since the thickness (30 μm) is smaller compared to the other dimensions, the actual applied pressure should be almost uniaxial along the *c*-axis, despite using a hydrostatic pressure technique. This relationship between sample configuration and anisotropic pressure has demonstrated previously in ref.20. The resistivity measurements were performed under magnetic fields (0.1, 0.2, 0.4, 0.6, 0.8, 1, 2, 3, 4, 5, 6 and 7 T) with directions both parallel to the *ab* plane (*H*//*ab*) and *c* axis (*H*//*c*), at the ambient pressure and 2.3 GPa.

Results and discussion

Figure 1 shows the XRD diffraction pattern for $PrO_{0.5}F_{0.5}BiS_2$. All diffraction peaks are assigned to the (00l) indices of $PrO_{0.5}F_{0.5}BiS_2$. The *c* lattice parameter is estimated to be 1.3366(9) nm from peak positions. This value is consistent with reported result of polycrystalline $PrO_{0.5}F_{0.5}BiS_2$ (*c* = 1.3362(4) nm) [5]. The inset of figure 1 displays the SEM image of the sample configuration for a high pressure measurement. From the EDX results, elemental ratio is estimated as following (Pr: O: F: Bi: S = 0.98: 0.63: 0.5: 0.93: 1.96). This is almost the same as the stoichiometric composition of $PrO_{0.5}F_{0.5}BiS_2$.

In this study, $T_c^{onset}$ was regarded as the cross point of the fitting lines in normal conducting state and in the drop area during the transition. $T_c^{zero}$ was also regarded as the cross point of the line for zero resistivity and the $\rho$-*T* curve. Temperature dependence of resistivity for $PrO_{0.5}F_{0.5}BiS_2$ under various pressures is shown in figure 2, the $T_c^{onset}$ is observed at 4.3 K under ambient pressure, and the behavior of resistivity shows a shallow valley at around 200 K. When increasing pressure to 0.7 GPa, the $T_c^{onset}$ suddenly increases up to 8.56 K and the shallow valley shifts to lower temperature around 170 K. At 1.5 GPa, the highest $T_c^{onset}$ and $T_c^{zero}$ are obtained at 8.72 K and 8.23 K respectively. The resistivity in the normal conducting state also changes from semiconducting-like behavior to metallic behavior from 0.7 to 1.5 GPa. With further increase in pressure, $T_c^{onset}$ and $T_c^{zero}$ gradually decrease down to 8.47 K and 7.87 K. The metallic behavior is still observed and the slopes of the resistivity are almost comparable between 1.5 GPa and 2.3 GPa. After releasing pressure, a $\rho$-$T$ curve again shows the shallow valley, and $T_c$ also changes to lower phase. Therefore, it is found that the phase transition reversibly occurs in accordance with applied pressure.

As shown in the left inset of figure 2, a step in resistivity is observed under 0.7 GPa near $T_c^{zero}$. This step like behavior of resistivity is the one of the pieces of evidence for the coexistence of low-$T_c$ and high-$T_c$ phases. Above 1.5 GPa, the second drop at around 5 K completely disappears, and the superconducting transition becomes quite sharp. The transition width is only around 0.5 K. It is suggested that the low-$T_c$ phase completely converts into the high-$T_c$ phase. However, in the case of polycrystalline $PrO_{0.5}F_{0.5}BiS_2$, the high-$T_c$ phase has not been observed under hydrostatic pressure, above even 2 GPa [18]. This means that uniaxial pressure along the $c$-axis effectively induces the high-$T_c$ phase.

As mentioned above, it has been reported that the phase transition between low and high-$T_c$ phases originates from a structural transition from tetragonal to monoclinic structure, which is induced by sliding between the two $BiS_2$-layers along the a-axis [16]. Assuming this structural transition, it is suggested that uniaxial pressure along c-axis should be more effective to induce this sliding along a-axis than hydrostatic pressure. Actually, in this study, we successfully obtained the high-$T_c$ phase even at 0.7 GPa.

To estimate the $H_{c2}$ and the anisotropy in both the low- and high-$T_c$ phases, the temperature dependences of resistivity under magnetic field with along $c$-axis and $ab$ plane were measured as shown in figure 3 (a)-(d). When the magnetic field is parallel to the $c$ axis, $T_c$ drastically decreases with increasing magnetic field. This is typical behavior observed in superconductors with a layered structure. However, in figure 3(d),

$\rho$-$T$ curve under magnetic field shows some strange behavior. A hump suddenly appears below $T_c^{onset}$, when a magnetic field of even 0.1 T is applied. Furthermore, this hump becomes larger and shifts to lower temperature with increasing magnetic field. On the other hand, the hump cannot be observed at all under $H//ab$. At this present stage, the origin of the hump cannot be explained. It could be an anisotropic phenomenon concerned with the magnetic field along the $c$ axis under high pressure and should be addressed as future work. Also, it is difficult to decide the $T_c^{onset}$ from these $\rho$-$T$ curves. Therefore, the top of the hump is adopted as the $T_c^{onset}$ in this study.

Figure 4 shows the temperature dependence of the upper critical field ($H_{c2}$). 85 % of resistivity at $T_c^{onset}$ is adopted as a criterion for the estimation of $H_{c2}$. In addition, $H_{c2}$ obtained from magnetic field $H//ab$ and $H//c$ are described as $H_{c2}^{//ab}$ and $H_{c2}^{//c}$ respectively. Anisotropy ($\gamma$) is also estimated from the following formula: $\gamma = H_{c2}^{//ab} / H_{c2}^{//c}$. The superconducting parameters at 0 K ($H_{c2}^{//ab}$ (0), $H_{c2}^{//c}$ (0), $\xi_c$ (0), $\xi_{ab}$ (0) and $\gamma$ (0)) are summarized in table 1, where $\xi_c$ (0) and $\xi_{ab}$ (0) are the coherence length along the $c$ axis and the $ab$-plane, respectively. They are obtained by the Ginzburg-Landau relation of following formula. $H_{c2}^{//c} = \Phi_0/2\pi\xi_{ab}^2(0)$ and $H_{c2}^{//ab} = \Phi_0/2\pi\xi_{ab}(0)\xi_c(0)$, where $\Phi_0$ is flux quantum. When changing from a low-$T_c$ phase to a high-$T_c$ phase, the $\gamma$ decreases from 20 to 9.3 and $\xi_c$ increases from 0.56 to 0.71 nm. These findings indicate the enhancement of three-dimensionality in the superconducting state, and it is thought to be caused by shrinkage in the $c$ lattice constant by applied uniaxial pressure along $c$ axis. The $c$ lattice constant of $PrO_{0.5}F_{0.5}BiS_2$ is estimated to be 1.3366 nm at ambient pressure in this research. This value should be shorter when applying pressure. In addition, obtained $\xi_c$ in the high-$T_c$ phases is 0.71 nm. This length is comparable to the height of the blocking layer of $PrO_{0.5}F_{0.5}BiS_2$. Therefore, it is suggested that even a small change in the $c$ lattice constant is effective in enhancing the overlap of the Cooper pair wave function along the $c$ axis, which suppress the anisotropic superconductivity.

Moreover, the observed metallic behavior of the high-$T_c$ phase in the normal conducting state as shown in figure 2 would be caused by the enhancement of three-dimensional conductivity due to a shrinkage in the $c$ lattice constant.

# Conclusion

We successfully obtained a high-$T_c$ phase of $PrO_{0.5}F_{0.5}BiS_2$ by applying pressure, and found that uniaxial pressure along c-axis effectively induces this high-$T_c$ phase. The maximum $T_c^{onset}$ and $T_c^{zero}$ are obtained at 8.72 K and 8.23 K respectively at 1.5 GPa. When the low-$T_c$ phase completely changes to the high-$T_c$ phase, the resistivity also changes from semiconducting-like behavior to metallic behavior. Furthermore, this transition between the low- and high-$T_c$ phases reversibly occurs in accordance with the applied pressure. This material shows anisotropic superconductivity. However, the phase change to a higher $T_c$ suppresses this anisotropic property. The $\gamma$ decreases from 20 to 9.3 and $\xi_c$ increases from 0.56nm to 0.71 nm. These findings indicate the enhancement of three-dimensionality in the superconducting state, which is induced by a shrinkage in the $c$ lattice constant due to uniaxial pressure along the $c$-axis.

# Acknowledgment

This work was supported in part by the Japan Science and Technology Agency through Strategic International Collaborative Research Program (SICORP-EU-Japan) and Advanced Low Carbon Technology R&D Program (ALCA) of the Japan Science and Technology Agency.

Caption

Fig. 1

(Color online) XRD patterns of $PrO_{0.5}F_{0.5}BiS_2$. All diffraction peaks are assigned to the (00l) indices. Inset shows SEM image of the sample configuration for a high pressure measurement.

Fig. 2

(Color online) Resistivity versus temperature from 300 K to 2 K for $PrO_{0.5}F_{0.5}BiS_2$ under varying pressure. Left inset shows an expanded view around $T_c^{zero}$. Black arrow shows the second drop caused by low-$T_c$ phase. Right inset shows an expanded view around $T_c^{onset}$.

Fig. 3

(Color online) Resistivity versus temperature for the low-$T_c$ and high-$T_c$ phases under magnetic field with the different directions, which are parallel with $ab$ plane ($H//ab$) and $c$ axis ($H//c$) respectively.

Fig.4

(Color online) Upper magnetic field versus temperature for the low-$T_c$ (black) and high-$T_c$ (red) phases. Solid circles denote the $H_{c2}^{//ab}$. Open circles denote $H_{c2}^{//c}$. The dashed lines denote the straight-line approximation.

Table 1. Superconducting parameters for $Pr(O_{0.5}F_{0.5})BiS_2$.

| Phase | $dH_{c2}^{//ab}/dT$ (T/K) | $dH_{c2}^{//c}/dT$ (T/K) | $H_{c2}^{//ab}(0)$ (T) | $H_{c2}^{//c}(0)$ (T) | $\xi_{ab}(0)$ (nm) | $\xi_c(0)$ (nm) | $\gamma(0)$ |
|---|---|---|---|---|---|---|---|
| Low $T_c$ | -11.35 | -0.58 | 52 | 2.6 | 11.2 | 0.56 | 20 |
| High $T_c$ | -8.37 | -1.16 | 70 | 7.5 | 6.62 | 0.71 | 9.3 |

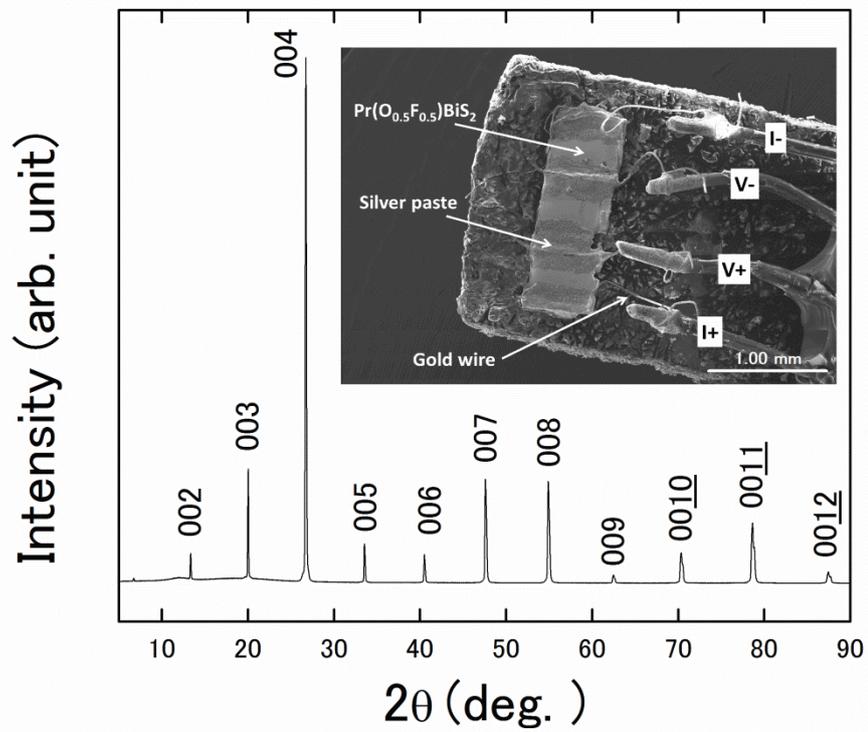

Figure 1

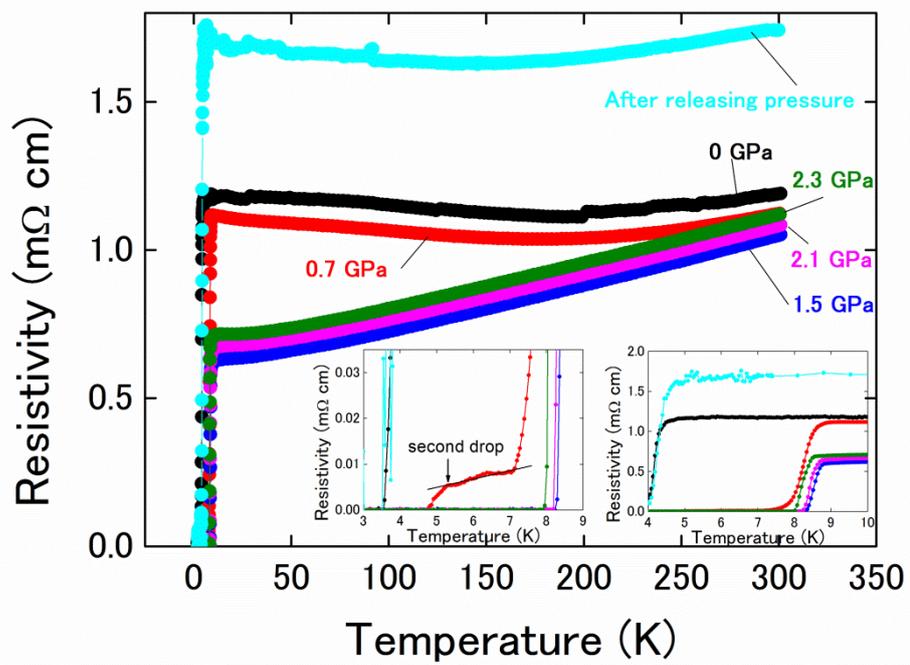

Figure 2

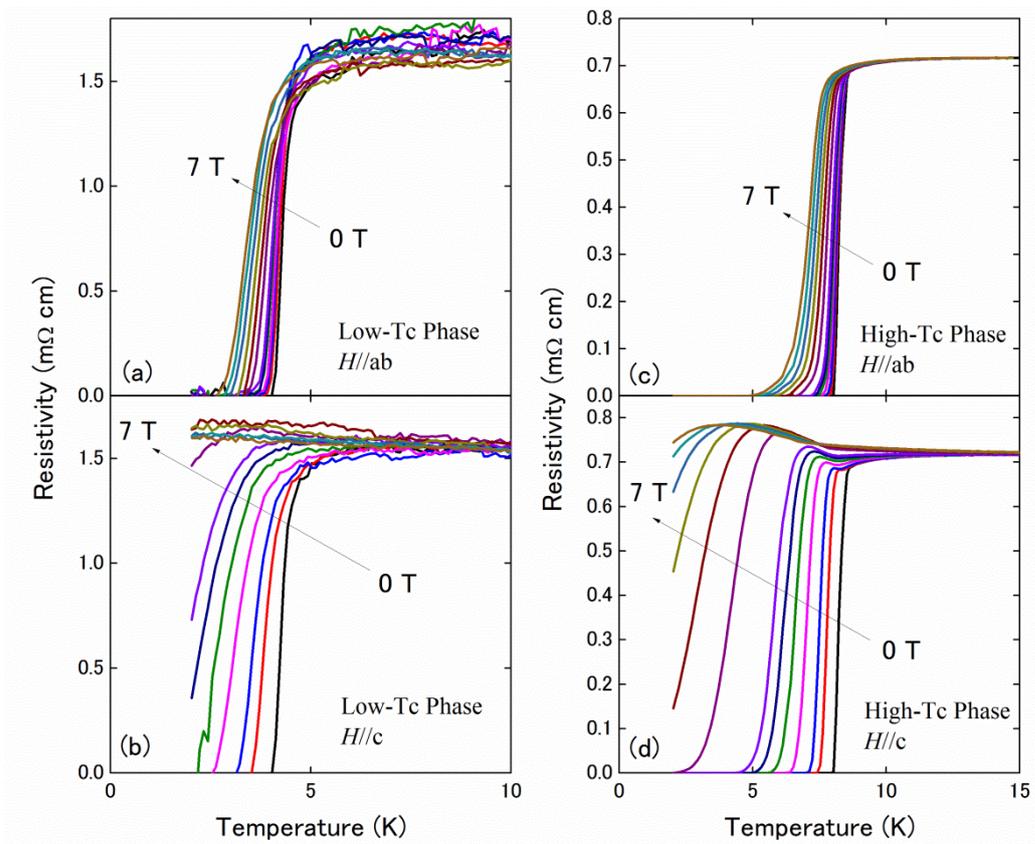

Figure 3

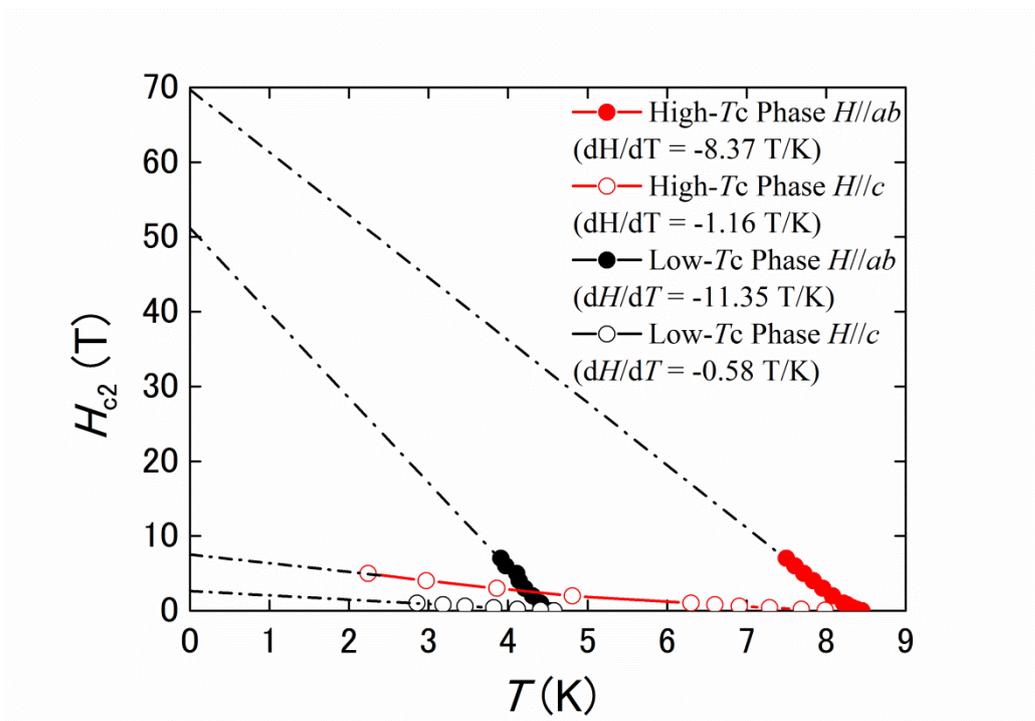

Figure 4